\documentclass{article}

\usepackage[]{graphicx}

\newcommand{\be}{\begin{equation}}
\newcommand{\ee}{\end{equation}}

\begin{document}

% Redefine "plain" pagestyle
%\makeatletter	   % `@' is now a normal "letter' for LaTeX
%\renewcommand{\ps@plain}{%
%     \renewcommand{\@oddhead}{\textrm{Ari Brynjolfsson}\hfil\textrm{\thepage}}% 
%     \renewcommand{\@evenhead}{\@oddhead}%
%     \renewcommand{\@oddfoot}{}% empty footer
%     \renewcommand{\@evenfoot}{\@oddfoot}}
%\makeatother }    % `@' is restored as a "non-letter" character

\title{Plasma Redshift, Time Dilation, and Supernovas Ia}         % Enter your title between curly braces
\author{Ari Brynjolfsson \footnote{Corresponding author: aribrynjolfsson@comcast.net}}

        % Enter your name between curly braces
\date{\centering{Applied Radiation Industries, 7 Bridle Path, Wayland, MA 01778, USA}}          % Enter your date or \today between curly braces

\maketitle
%Abstract

\begin{abstract}  The measurements of the absolute magnitudes and redshifts of supernovas Ia show that conventional physics, which includes plasma redshift, fully explains the observed magnitude-redshift relation of the supernovas.  The only parameter that is required is the Hubble constant, which in principle can be measured independently.  The contemporary theory of the expansion of the universe (Big Bang) requires in addition to the Hubble constant several adjustable parameters, such as an initial explosion, the dark matter parameter, and a time adjustable dark energy parameter for explaining the supernova Ia data.  The contemporary Big Bang theory also requires time dilation of distant events as an inherent premise.  The contention is usually that the light curves of distant supernovas show or even prove the time dilation.  In the present article, we challenge this assertion.  We document and show that the previously reported data in fact indicate that there is no time dilation.  The data reported by Riess et al. in the Astrophysical Journal in June 2004 confirm the plasma redshift, the absence of time dilation, dark matter, and dark energy.

\end{abstract}



%Table of Contents

% Redefine "plain" pagestyle
\makeatletter	   % `@' is now a normal "letter' for LaTeX
\renewcommand{\ps@plain}{
     \renewcommand{\@oddhead}{\textit{Ari Brynjolfsson: Time Dilation and Supernovas Ia}\hfil\textrm{\thepage}}% 
     \renewcommand{\@evenhead}{\@oddhead}
     \renewcommand{\@oddfoot}{}% empty footer
     \renewcommand{\@evenfoot}{\@oddfoot}}
\makeatother     % `@' is restored as a "non-letter" character

% Set to use the "plain" pagestyle
\pagestyle{plain}

%Section 1: Introduction

\section{Introduction}

The remarkable measurements of the supernovas' absolute magnitudes and redshifts by the many well-equipped groups of experienced researchers help us define some important cosmological parameters.  The good quality of this work gives us opportunity to test the different cosmological models.  Analyses of the light curves of low redshift supernovas Ia show that the width of the light curves (the increase and subsequent reduction in the light intensity with time) varies with the maximum absolute magnitudes of the supernovas.  The increased width of the light curves with the increasing brightness of the supernovas is reasonable, as we usually expect a larger explosion to result in bigger dimensions of the explosion, and therefore longer time for explosion to expand, and longer time for the larger amount of energy to decay after maximum.  The increased width, however, can be partially obscured by the usually assumed time-dilation effect.  The observed width of the light curves of the high-redshift supernovas is reduced by dividing the width by the assumed time-dilation factor, $(1+z),$ before it is compared with the template curves for the nearby supernovas, where the time dilation is insignificant.  This reduced width of the light curve for a distant supernova then results in a reduced brightness estimate.

\indent   When applying the plasma-redshift theory [1] (see in particular Eqs.\,(54)-(56) and Table 4 of that source), there is no time dilation.  Therefore, had we applied the plasma redshift when interpreting the observations, the supernovas would be estimated to be brighter, ${\Delta} {\rm{M}} = -2.5 \,{\rm{log}}\,(1+z),$  than the estimate obtained assuming time dilations.  However, the concurrent dimming, ${\Delta} {\rm{M}} = 5 \,{\rm{log}}\,(1+z),$ caused by the Doppler shift on the intergalactic plasma electrons, in accordance with the plasma-redshift theory, causes a dimming that is twice the dimming caused by the time dilation.  This additional dimming is however reduced, because the distance modulus is also different in the two theories.  In fact the dimming with increasing $z$ of the supernovas in the plasma-redshift theory is almost equal to the dimming with increasing $z$ of the supernovas in the contemporary Big Bang theory.  Occasionally, the researchers felt incorrectly that they had proven the time-dilation effect, because they were unaware of the plasma-redshift theory.

\indent  In Section 2, we will first explore what the observations of the supernovas indicate about the time dilation.  As we will see, the data, contrary to common beliefs, strongly indicate that there is, in fact, no time-dilation effect.  The observed lack of time-dilation effect contradicts the contemporary expansion or Big Bang theory and indicates thus that we should explore other theories for explaining the cosmological perspectives.  In section 3, we find that the supernova observations nicely confirm the redshift magnitude relation predicted by the plasma-redshift theory, which has no time dilation.

%Section 2: What do the supernova observations indicate about the time dilation?

\section{What do the supernova observations indicate about the time dilation?}

The research teams investigating the absolute magnitudes, ${\rm{M}}_{max}$, of the nearby supernovas often use the rate of decrease in brightness or rate of increase in absolute magnitude, $\Delta{{\rm{M}}_{15}},$ during the first 15 days after the maximum brightness, as a measure of the width of the light curve.  The greater the value of $\Delta{{\rm{M}}_{15}},$ the steeper is the decay of the light intensity, or the smaller is the width, $w,$ of the light curve.  The increase in maximum magnitude (corresponding to a decrease in the maximum brightness), is then also found to increase roughly proportional to the change in magnitude $\Delta {\rm{M}_{15}} .~$ For the low-redshift supernovas, this relation appears consistent and reliable.  

\indent  The researchers also use the width, $w = s(1+z),$ of the supernova's light curve, where $s$ is called a stretch factor, and $(1+z)$ is the time-dilation factor.  For the nearby supernovas, we have that the width is roughly proportional to $1/\Delta{{\rm{M}}_{15}}.~$  When we observe low-redshift supernovas, the width, $w,$ of the light curve is found to increase with the brightness.  For low-redshift supernovas $( z \leq 0.1 )$ , the change in the factor of $(1+z)$ is small, so it is principally the stretch factor, $s,$ that increases when the width increases and the brightness increases (or the absolute magnitude decreases).  When observing a distant supernova at a redshift of $z > 0.1,$ the supernova researchers will divide that measured light-curve width, $w,$ by a time-dilation factor, $(1+z),$ to obtain an estimate of the stretch factor, $s ,$ representing the width of a corresponding supernova at close range.  From this reduced width, they estimate the maximum intensity of the distant supernova, based on the maximum brightness or magnitude for corresponding width of the nearby supernovas.

\indent  When evaluating distant objects, there is always a tendency for a Malmquist bias.  Amongst the distant supernovas, we are likely to observe the brightest members of that group of supernovas.  The question then arises, how big is the Malmquist bias?  It is noteworthy, that in sample of 10 distant supernovas reported in table 6 of the article by Riess et al. [2] the two supernovas with the highest redshift have the greatest widths of their light curves, while the five supernovas with the smallest redshift have the smallest widths, even after the division by the time-dilation factor $(1+z) .~$  This indicates a slight Malmquist bias.

\indent  For nearby $(z \leq 0.1)$ supernovas, which are practically independent of the time-dilation effect, Phillips [3] (see in particular Table 2 of that source) finds that if we write peak brightness magnitude as:
\be
{\rm{M}_{max}} = {a} + {b} \,\Delta {\rm{M}}_{15}
\ee
\noindent  then the experiments for nine low-redshift sample of supernovas, with $z \leq 0.01 ,$ indicate that for the spectral bands B, V, and I, the values of a and b are respectively as shown in the following array of equations:
\be
\left\{ {\begin{array}{*{20}c}
   \rm{B}...: & {^{\rm{B}}{\rm{M}_{max}}\,=\,} & \!\!\!-21.726\,(0.498) & \!\!\!+ ~ 2.698\,(0.359)\,\Delta{{\rm{M}}_{15}}, & \quad ~~\sigma (^{B}{\rm{M}_{max}}) = 0.36  \\
   \rm{V}...: & {^{\rm{V}}{\rm{M}_{max}}\,=\,} & \!\!\!-20.883\,(0.417) & \!\!\!+ ~ 1.949\,(0.292)\,\Delta{{\rm{M}}_{15}}, & \quad ~~\sigma (^{V}{\rm{M}_{max}}) = 0.28  \\
   \rm{I}...: & {^{\rm{I}}{\rm{M}_{max}}\,=\,} & \!\!\!-19.591\,(0.415) & \!\!\!+ ~ 1.076\,(0.273)\,\Delta{{\rm{M}}_{15}}, & \quad ~~\sigma (^{I}{\rm{M}_{max}}) = 0.38  
\end{array}} \right\}. 
\ee
\noindent  The effect of possible time dilation on the magnitude estimates of these low redshift supernovas is less than 0.0017 redshift units, and therefore insignificant.

\indent  Phillips [3] (see last sentence of the abstract of that source) makes the following comment: "Considerable care must be exercised in employing Type Ia supernovae as cosmological standard candles, particularly at large redshifts where Malmquist bias could be an important effect".  

\indent  Subsequently, Phillips et al. [4] slightly modified the approach by using second order polynomials in $(\Delta {\rm{m}_{15}}(\rm{B}) - 1.1)$ instead of the linear equations above, and by taking into account the absorption in our Galaxy and the host galaxy, and by considering a larger number of supernovas (62 instead of 9).  Phillips et al. [4] comment (see the introduction of that source) that: "... there is now abundant evidence for the existence of a significant dispersion in the peak luminosities of these events at optical wavelengths, the absolute magnitudes fortuitously appear to be closely correlated with the decay time of the light curve..."  

\indent  Goldhaber et al. [5] adjust the R-band photometric data to one maximum intensity $I_{max}$ and scale the time axis by the light-curve width, $w.$  This width factor has the form $w \equiv s\,(1+z),$ where $s$ is the stretch factor, and $(1+z)$ is the time-dilation factor.  For a sample of 35 supernovas out of total of 42 their Table 1 shows that the stretch factor $s$ varies from between 0.71 and 1.55, or about a factor of 2.18, while the $(1+z)$-values vary from 1.172 to 1.657, or a factor of 1.41.  The 7 supernovas that were excluded would, if included, not have changed their conclusions

\indent Goldhaber et al. [5] analyzed similarly 18 of the 29 SNe of the Calan/Tololo set.  Their Table 2 shows that the stretch factor $s$ varies from 0.53 to 1.12, or about a factor of 2.11, while the $(1+z)$-values varied from 1.014 to 1.088, or a factor of 1.073.  The variations in the width are thus mostly due to variations in $ s .~$  Their Fig.\,1-(f) shows clearly that the width factor, w, results in a rather uniform light curve width as is to be expected.  This curve then also fits well to the template curve for Parab-18 as demonstrated in their Fig.\,2-(a).

\indent  Goldhaber et al. [5] show in their Fig.\,3-(a) for a sample of 42 high-redshift supernovas that the light curve width, $w$ is proportional to ${(1+z)}.~$  However, this is self-evident, because according to their definition $w \equiv s\,(1+z).~$  They could similarly have shown that the width, $w,$ is proportional to $s .$  However, the relation between $s$ and $(1+z) \leq 1.83$ as shown in Fig.\,3-(b), which includes all 42 high-redshift supernovas, is meaningful and most important.  It indicates clearly that the variation in the average of $s$ with time-dilation factor $(1+z)$ is insignificant.  The variations in $s$ are thus independent of the time dilation, $(1+z).~$  This is surprising, because according to the contemporary expansion theory, we expect the Malmquist bias to result in an increase in the brightness and an increase in $s$ with increasing $z.$  It appears that the actual increase in the width, $s,$ and the brightness of the supernova is suppressed by the time dilation, which the supernova researchers use to reduce the width of the light curve, and thereby the brightness.   

\indent  By definition, the maximum light intensity is proportional to $10^{-0.4M_{max}}.~$  When we use the contemporary expansion theory, the light intensity increases roughly proportional to the light curve width, $w \equiv s\,(1+z).~$  According to Fig.\,3-(b) of [5], the width, $w ,$ appears to increase roughly proportional to $(1+z)$ with a large noise $s .~$  We can then write 
$ -0.4\,{\Delta} {{\rm{M}}_w} \approx {\rm{log}}\, w = {\rm{log}}\,(1+z) + {\rm{log}}\, {s} ,$ or

\be
 {\Delta} {{\rm{M}}_w} \approx - 2.5 \,{\rm{log}}\, w = - 2.5 \,{\rm{log}}\,(1+z) - 2.5 \,{\rm{log}}\, {s} .
\ee
The analysis by Goldhaber et al. [5] shows thus that the average value of $s$ is independent of $z.~$  Their analysis therefore indicates that the increase in light intensity in units magnitude is actually given by
\be 
{\Delta} {{\rm{M}}_z} = - 2.5 \,{\rm{log}}\,(1+z).~
\ee
\noindent  The supernova researchers, in accordance with the contemporary Big Bang theory, reduce the absolute magnitudes ${\rm{M}},$ when they use the time-dilation effect to reduce the width.  Had the they not corrected the magnitude by using the time dilation in their equations, the corresponding width factor would be $w = s' .~$  We would then get a reasonable increase in the brightness with $z.~$ The analysis by Goldhaber et al. [5] (see in particular their Fig.\,3-b of that source) thus indicates that there is no time dilation.  The observations, therefore, appear to contradict the contemporary Big Bang theory, which has time dilation as a basic premise.    

\indent  The following equation describes the major changes when we omit the time dilation:

\be
{\rm{M}} = {\rm{M}_{exp}} - 2.5\,{\rm{log}}\,(1+z)= {\rm{M}_1} + \Delta {\rm{M}_1} - 2.5\,{\rm{log}}\,(1+z)={\rm{M}_0} + {\Delta}{\rm{M}_0}
\ee
where
\[
{\rm{M}_0} ={\rm{M}_1}   
\]
and
\[
{\Delta}{\rm{M}_0} ={\Delta}{\rm{M}_1}-2.5\,{\rm{log}}\,(1+z)
\]

\noindent  In Eq.\,(5), the value of M is the absolute magnitude of the supernova without time dilation, while ${\rm{M}_{exp}} = {\rm{M}_1} + \Delta {\rm{M}_1}$ is the experimentally determined magnitude by the supernova researchers adhering to the Big Bang theory.  ${\rm{M}_1}$ is their estimated reference magnitude without the correction for light curve width, while ${\Delta}{\rm{M}}_1$ accounts for the correction for the light curve width.  The ${\rm{M}_0}$ and ${\Delta}{\rm{M}_0}$ are, respectively, the corresponding magnitude and magnitude correction, which the supernova researchers would have measured had they omitted the time dilation and used the plasma redshift to guide them.

\indent  When the researchers correct the observed values by omitting the time dilation, they would have to take into account the concurrent Doppler effect term  $+5\,{\rm{log}}(1+z).~$  In addition, the researchers would have to take into account the difference in the distance modulus in the plasma-redshift theory and in the contemporary Big Bang theory.

\indent  Brynjolfsson [1] has shown, (see in particular Eq.\,(56) and Table 4 of that source) that the magnitude redshift relation in the plasma-redshift theory is very similar to the corresponding relation in the contemporary Big Bang theory if we omit the acceleration and deceleration terms.  The omissions of acceleration and deceleration in the contemporary Big Bang theory are often not reasonable.  But at small redshifts, the similarity between the values derived from the two theories helps us understand why it is difficult for small $z$-values to see the difference between the plasma redshift theory and the contemporay expansion theory.

\indent  It is often misleading to refer to the plasma-redshift theory as a "tired light theory", because the plasma redshift leads to the dimming by the Doppler effect and to a different distance modulus from that in most "tired light theories".  

\indent  Omitting the time dilation when correcting the absolute magnitude, increases the brightness correction of the supernova by the absolute-magnitude correction-term $-2.5\,{\rm{log}}(1+z) .~$ The concurrent dimming caused by the Doppler effect on the electrons and the modification by the difference in the form of the distance modulus made it very difficult for the supernova researchers to discern these changes.  When applying the contemporary Big Bang theory, the reduction of the width of the light curve by the time dilation artificially reduces the Malmquist bias.  We will therefore back correct the magnitudes for the time-dilation effect, which in accordance with Eq.\,(5) changes the absolute magnitude from ${\rm{M_{exp}}}$ to ${\rm{M}} .~$  This change results in a small brightening of the observed supernovas with increasing $z,$ as is to be expected from the Malmquist bias.  This change is also in accordance with Fig.\,3-(b) of reference [5], which shows that the light-curve width-parameter $s$ is independent of the time dilation.  

\indent  In their Fig.\,4, Richardson et al. [6] show the distribution of the absolute magnitudes, ${\rm{M}}_{\rm{B}}$, at maximum intensity for 111 normal SNe Ia with a distance modulus $\mu \leq 40 .~$  The Malmquist bias for the brightest supernovas is not obvious, but if we omit the time-dilation effect it would be obvious.  The magnitudes when uncorrected for extinctions in the parent galaxies were compared with the magnitudes when corrected for the adopted extinctions of the galaxies.  When approximated with a gaussian distribution curve, the distribution moved slightly to the brighter side (as expected), from -19.16 to -19.46 , and the $\sigma$-value decreased from 0.76 to 0.56.  The brighter tail end of the measured distribution was slightly smaller than that of the gaussian distribution.  This possibly could indicate an upper limit.  In this context, we should take in to consideration, however, that if the time-dilation effects were removed, it would increase the brightness, especially, of the brighter tail end or at the higher $z$-values.

\indent  When we increase the redshifts from $(z \approx 0.01)$ to large redshifts $(z \approx 1)$, we increase the number of supernovas per redshift interval by a large factor.  The fact that observations are limited by the magnitude means that we can expect a significant Malmquist bias.  This is generally realized by the supernova researchers as exemplified by their frequent reference to the Malmquist bias.  However, because of their use of time dilation in the their estimates, the Malmquist bias was less obvious than expected.

%Section 3: Redshift magnitude relation

\section{Redshift magnitude relation}

In section 2 above, we have shown that the supernova data strongly indicate that there is no time dilation.  For a given light emission from a supernova, the omission of the time dilation increases the observed light intensity, or decreases the observed magnitude, M, of the supernova by a term ${\Delta} {\rm{M}} = - 2.5 \,{\rm{log}}\,(1+z).~$ 
For eliminating the time-dilation effect, we must in accordance with Eq.\,(5) change the experimentally determined value, $ {\rm{M_{exp}}}$, as determined by the supernova researchers to
\be
{\rm{M}}= {\rm{M_{exp}}} + {\Delta} {M} ={\rm{M_{exp}}} - 2.5 \,{\rm{log}}\,(1+z).~
\ee
\noindent  This new $ {\rm{M}}$ is then nearly free of the time-dilation effects.

\indent   When comparing the predictions of the plasma-redshift theory with experiments, we will use the $\rm{M}$-values, as determined by Eq.\,(6).  For ${\rm{M_{exp}}} ,$ we use the very good magnitude data reported by Riess et al. [7] (see in particular the expanded Table 5 of that source).  These $\rm{M}$-values, which have been corrected for the false time-dilation effect, can then be used for testing the different cosmological theories.

\indent  As shown by Brynjolfsson [1], the plasma redshift gives a very simple explanation of the magnitude-redshift relation for the observed supernovas.  This relation is (see Eq.\,(54) of reference [1])
\be
m - M = 5\,{\rm{log}}\,({\rm{ln}}(1+z)) + 7.5\,{\rm{log}}\,(1+z)  + 5\,{\rm{log}}\,(\frac{{10^{6} \, c}}{{H_0}}) - 5 + A_B . 
\ee
\noindent  where $ A_B $ is the absorptions of light from the supernova expressed in magnitude units, $c$ is the velocity of light in ${\rm{km\,s}}^{-1} ,$ and $H_0 $ is the Hubble constant in ${\rm{km\,s}}^{-1}\,{\rm{Mpc}}^{-1} .$  This equation has no adjustable parameters except the Hubble constant, which only moves the curve in Fig.\,1 up and down independent of $z.~$  The supernova experiments can be used to measure accurately the Hubble constant.  It is thus a very simple equation that must match the many experimental points. This equation contrasts the conventional magnitude redshift relation for the Big Bang cosmology, which in addition to the Hubble constant requires adjustable parameters for dark matter, and time dependant dark energy or a time dependant cosmological constant.  

\indent  In the reference [1] (see Fig.\,5 of that source), the plasma-redshift theory was compared with supernova data reported by Riess et al. [2].  Although the fit to the data is very good, it can be seen that the three supernovas with the largest redshift are slightly below the theoretically expected line in spite of the fact that the high z-data also pulled the theoretical curve down.  This is because the experimental data reported by Riess et al. [2] were not corrected for the false time-dilation effect conventionally used by the supernova researchers, who assumed the contemporary Big Bang theory when reporting their data.  The contemporary Big Bang theory has time dilation as an inherent premise.  Had we in Fig.\,5 of [1] applied the back correction for the time dilation given in Eq.\,(6) above, these three points with the largest redshift would have fallen on the theoretical curve predicted by the plasma redshift.

%Figure1

\begin{figure}[t]
\centering
\includegraphics[scale=.5]{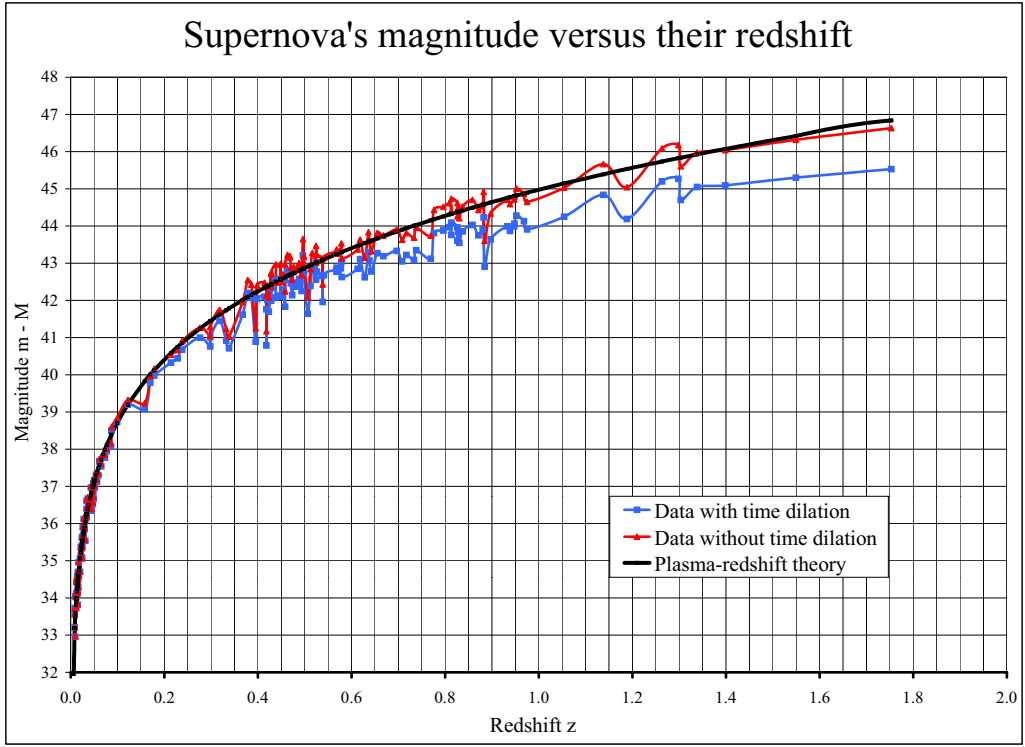}
\caption{The magnitudes, m-M, of supernovas on the ordinate versus their redshifts, $z$ from 0.0 to 2, on the abscissa.  The data include all 186 supernovas reported by Riess et al. [7] (see the expanded Tables 5 of that source).  The lower data points indicated with rectangles (blue) are as reported by Riess et al. [7] and include the time dilation, while the data points indicated with triangles (red) are free of time dilation.  The black curve shows the theoretical predictions of the plasma-redshift theory in accordance with Eq.\,(7), when the a Hubble constant of ${\rm{H}}_0 = 59.44 ~{\rm{km\,s}}^{-1}\,{\rm{Mpc}}^{-1}.$}
%\label{}
\vspace{2mm}
\end{figure}

\indent  With the new data reported by Riess et al. in Table 5 of reference [7], showing several supernovas in the range $1 \leq z \leq 1.755 ,$ it became necessary to eliminate the false time-dilation effect in accordance with Eq.\,(6) above.  With this correction, we can use all the newly reported supernova data listed in Tables 5, by Riess et al. [7] and check them against Eq.\,(7).  In Fig.\,1, we have used all of their 186 data points, the samples indicated with gold as well as silver. 

\indent  Three correction methods for the effect of light-curve widths and shapes have been used to determine the magnitude of some of the supernovas.  For 10 of the supernovas the magnitude was determined using the MLCS-method and the $\Delta {\rm{M}_{15}}(\rm{B})$-method and reported in Tables 5 and 6 of reference [2], and using the MLCS2k2-method in Table 5 of reference [7].  These three methods for determination of ${\rm{M}_{exp}},$ when compared with the predictions of the plasma-redshift theory appeared to be about equally good.  The MLC method scored slightly better than the other two; but because of the small sample, the difference was not significant.  We use therefore the data as reported by Riess et al. [7].

\indent  The plasma redshift predicts a small additional redshift due to the corona of the Milky Way Galaxy and the corona of the host galaxy.  For this reason, we have reduced all the redshifts by an amount $\Delta z = -0.00185.~$  This is nearly an insignificant correction, but in principle a correction on this order of magnitude should be applied when using the plasma-redshift theory.  This corresponds to an average redshift of $\Delta z = -0.000925~$ for each galaxy.  The corresponding Hubble constant is ${\rm{H}}_0 = 59.44 ~{\rm{km\,s}}^{-1}\,{\rm{Mpc}}^{-1}.~$

\indent  Riess et al. [7] characterize 157 out of the 186 SN Ia (in their Table 5) as gold samples, while the remaining 29 are silver samples.  When we compare the corrected observed magnitudes with the predictions of the plasma-redshift theory, we find that four of the supernovas (SN1997as, Sn1998I, SN200ea, and Sn2001iv) had magnitudes in the range $[({\rm{m}}-{\rm{M}})-({\rm{m}}-{\rm{M})_{pr}}]\leq -0.8,$ where the subscript 'pr' referes to the plasma-redshift theory.  Of these 3 are gold samples and 1 silver sample.  All of these four samples had negative values for the quantity inside the brackets, which indicates that the large and one sided deviations from the predicted values were possibly due to a larger absorption $A_B$ than those assumed for these four supernovas.  Analogously, we find that six of the supernovas are in the interval $-0.8 \leq [({\rm{m}}-{\rm{M}})-({\rm{m}}-{\rm{M})_{pr}}]\leq -0.5.~$  Of these 3 are gold samples and 3 silver samples.  Similarly, we find that four of the supernovas are in the interval $+0.8 \geq [({\rm{m}}-{\rm{M}})-({\rm{m}}-{\rm{M})_{pr}}]\geq +0.5.~$  Of these 2 are gold samples and 2 silver samples.  These last mentioned, 6 and 4, supernovas, are within the expectation of a gaussian distribution for this large number of supernovas. 

\indent  In spite of the relatively large one sided deviations from the theoretical curve for four of the samples, the distribution is nearly gaussian with a standard deviation for an individual sample of about $\sigma _{\rm{M}}= 0.30$ of a magnitude.  The slightly skewed distribution with heavier negative tail indicates an additional absorption in some of the supernovas.  The standard deviation in the average value of the absolute magnitude determining the theoretical curve is only $\sigma _{\rm{M}_{av}}= 0.022$ magnitude.  According to Eq.\,(7), the value of the Hubble constant is then ${\rm{H}}_0 = 59.44 \pm 0.6 ~{\rm{km\,s}}^{-1}\,{\rm{Mpc}}^{-1}.~$  This high accuracy in the determination of ${\rm{H}}_0$ applies only to the internal consistency between the plasma-redshift theory and the measurements of the supernovas and does not apply to the actual uncertainties, which depend on the uncertainties in determining the absolute magnitudes of the supernovas at a well determined absolute distances.

\indent  In conclusion, the data from the supernovas Ia indicate that there is no time dilation.  As Fig. 1 shows, the data support with very high accuracy the plasma redshift theory, which has no time dilation.  The plasma redshift theory rejects the data with time dilation with high degree of confidence.

\indent  The plasma-redshift cross-section [1] follows directly from well-proven conventional basic physics.  In addition to explaining the magnitude-redshift relation for the supernovas Ia, the plasma redshift helps explain the heating of the solar corona, the galactic corona, the heating of intergalactic plasma, and the cosmic microwave background [1].  The plasma redshift, when combined with the solar redshift experiments, leads to weightlessness of photons in a local system of reference (and repulsion of photons in a gravitational field when observed by distant observer).  This fact invalidates the equivalence principle [1].  All the many experiments that incorrectly have been assumed to prove the equivalence principle are in the domain of classical physics, and therefore do not make it possible to detect quantum mechanical effects, which are essential for observing the weightlessness of photons.  Only the solar redshift experiments are in the domain of quantum mechanics, and these show clearly the repulsion of photons [1], and therefore that the equivalence principle is false.  This in turn can lead to quasi-static universe without Einstein's $\Lambda$-coefficient [1].  The plasma-redshift explanations have no need for dark matter, dark energy, nor black holes.

%References

\end{document}